# A New Backscattering Dual-Polarized Rectenna for Wireless Power Transfer and IoT Applications

Taki E. Djidjekh[#1], Quentin Bernyer[#2], Alexandru Takacs[*3]
[#]LAAS-CNRS, Université de Toulouse, 31400 Toulouse, France
[*]LAAS-CNRS, UPS, Université de Toulouse, 31400 Toulouse, France
{[1]taki.djidjekh, [2]quentin.bernyer, [3]alexandru.takacs}@laas.fr

***Abstract*— This paper proposes an innovative dual-polarized backscattering rectenna that operates in two distinct modes—energy harvesting and backscattering modulation—driven by two-bit digital control signals. By utilizing two orthogonal (co- and cross-) polarizations, the design represents a versatile candidate for IoT applications such as battery-free wireless sensing, identification, localization, and communication. The rectenna's dual functionality is validated through its integration into a proof-of-concept battery-free wireless sensor, where it operates both as an energy harvester and as a dual-polarized backscattering modulator. As a proof of concept, a 16-byte AES-128 encrypted payload is backscattered over the wireless power transfer link to enhance the resilience of a battery-free Bluetooth Low Energy (BLE) wireless sensor against replay, relay, and eavesdropping attacks.**

## I. Introduction

In the context of IoT applications, the use of Battery-free Wireless Sensors (BWS) [1]–[4] powered at distance by Wireless Power Transfer (WPT) [5] or by Energy Harvesting (EH) [6] is a very promising technology. In conjunction with these technologies, the use of backscattering is also an energy-efficient and cost-effective solution (as no dedicated transmitter is required at the tag level) for identification (e.g., RFID), localization (e.g., Radar), sensing, and communication.

A key element in building an efficient BWS based on WPT or electromagnetic EH is the rectenna, which is composed of an antenna and an RF rectifier that acts as an energy harvesting device, converting the incoming electromagnetic power into DC power. For backscattering-based applications, the backscattered signal is generated by modulating the antenna load, which creates a modulated reflection of the incoming waveform received by the same antenna. Typically, the information encoded in the backscattered waveform is used for identification, localization, sensing, or communication purposes.

This paper focuses on a new and original topology for a Backscattering Rectenna (BR) able to operate in two modes: energy harvesting and backscattering modulation. As a backscattering modulator, this BR generates an orthogonal/cross-polarized waveform with respect to the incoming WPT waveform. As for polarimetric radar application [7], this polarization separation mitigates clutter and reduces cross-jamming effects in real environments. A backscattering rectifier was recently proposed in [8], implemented using a MOSFET incorporated into the rectifier topology. By controlling the gate of this transistor, it is possible to switch between the harvesting and backscattering modes and to control the load of the antenna/rectenna. The backscattered waveform in [8] has the same polarization as the incoming WPT waveform. In this paper, a more versatile solution is proposed, which consists of using an RF fail-safe switch connected to two cross-polarized antennas and with a backscattering rectifier. This 2-bit BR is very versatile and can generate as function of logical selection various backscattered waveforms with two orthogonal polarizations by modulating the switch between the two orthogonally polarized antennas or the MOSFET's gate of the backscattering rectifier, and consequently between the three operating modes: energy harvesting and backscattering with co- and cross-polarization.

Section II presents the architecture of the 2-bits backscattering rectenna topology while Section III reports experimental results for a BLE-based BWS with additional feedback channel implemented by using the BR. Section IV provides discussion and concluding remarks.

## II. Backscattering Rectenna Topology and Characterizations

The architecture of the proposed 2-bit backscattering rectenna integrated into a BWS is represented in Fig. 1a).

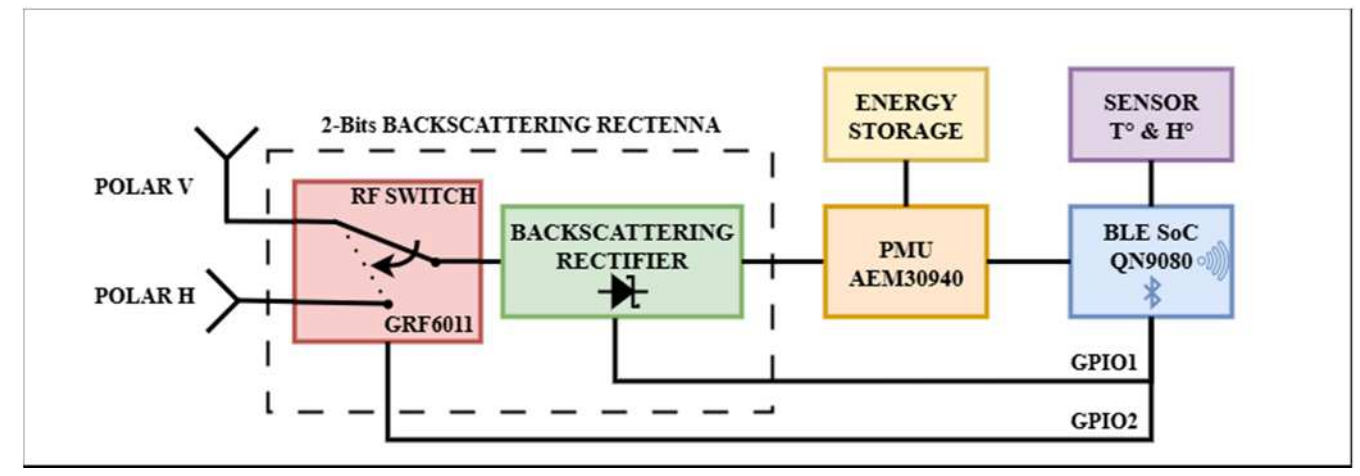

(a)

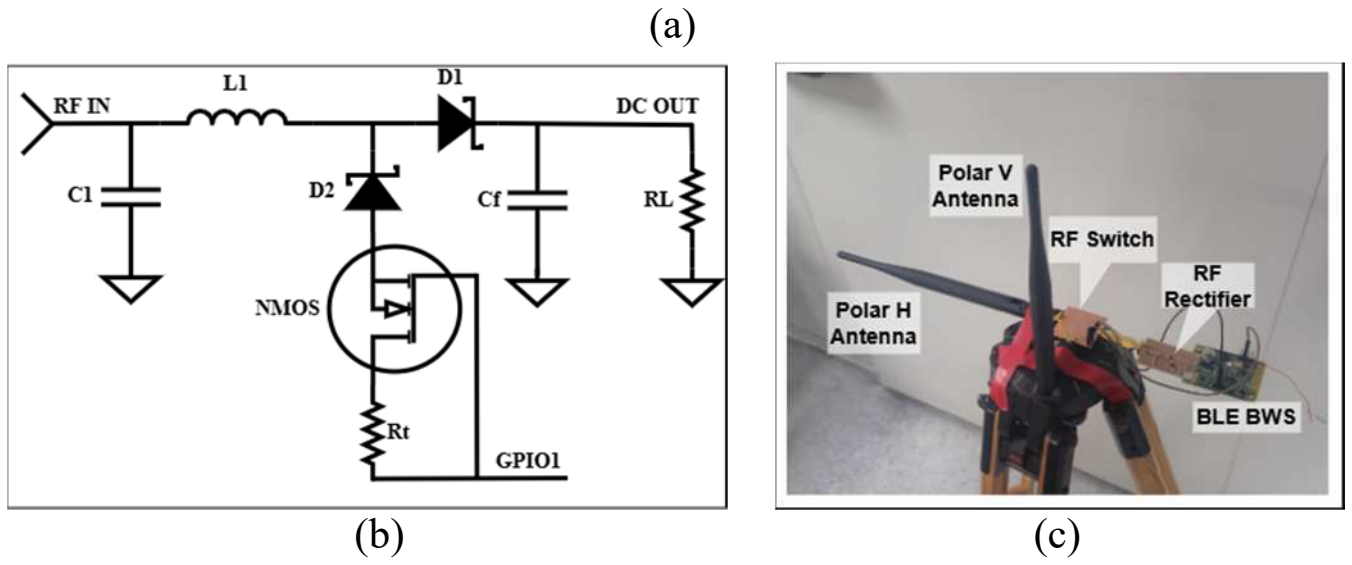

(b) (c)

Fig. 1. (a) Architecture of the battery-free wireless sensor with a 2-bit backscattering rectenna; (b) The schematic of the backscattering rectifier circuit; (c) Photograph of the fabricated prototype.

Fig. 1b) shows the schematic of the backscattering rectifier circuit: $L_1$ = 27 nH, $C_1$ = 3.3 pF, $D_1$–$D_2$ (SMS7630-005LF, Skyworks), the N-Channel MOSFET (BSS123, ON Semiconductor), and $C_f$=100 pF. The rectifier is impedance-matched to the antenna/RF input when $GPIO_1$ = 0 V (MOSFET Off), and intentionally mismatched when GPIO1 = 3.3 V (MOSFET On) to enable backscatter modulation.

The proposed BR consists of two cross-polarized antennas (vertical, polar-V, and horizontal, polar-H), an RF backscattering rectifier operating in the 868 MHz ISM band, and an RF fail-safe SPDT switch (GRF6011, Guerrilla RF). Two logic control signals ($GPIO_1$ and $GPIO_2$) configure the BR operating mode and generate the backscattered waveform through On-Off Keying (OOK) modulation.

The fail-safe switch maintains a default state without any supply voltage, which is well suited for energy harvesting using the polar-V antenna. The proposed topology is reconfigurable (as depicted in Fig. 1 and summarized in Table 1), enabling a combined energy harvester/backscattering modulator that operates in three distinct modes:

Table 1. Backscattering Rectenna operation modes.

| $GPIO_1$ | $GPIO_2$ | Mode | Function |
|---|---|---|---|
| 0 | 0 | Energy Harvesting | Providing the dc energy to the PMU |
| 0 | Toggling | Backscattering mode (Co-polarization) | OOK modulator in backscattering mode |
| Toggling | 0 | Backscattering mode (Cross-polarization) | OOK modulator in backscattering mode |
| 1 | 1 | Not used | None |

(i) Energy Harvesting – In this mode, both $GPIO_1$ and $GPIO_2$ are OFF (0V), the incoming WPT energy received by polar V antenna is directed by the RF switch (that is in failsafe state) to the RF rectifier to generate the DC energy. In a typically BWS topology this energy is stored via a Power Management Unit (PMU) into the storage capacitor. In this mode no DC energy is consumed by the RF switch but approximately 0.4 dB insertion losses are added.

(ii) Backscattering Modulator (Cross-Polarization) – In this mode, the $GPIO_2$ is activated (toggling between 0V and 3.3V) and the incoming WPT signal received by the polar V antenna is directed to polar H antenna for generating the backscatter waveform. The MOSFET transistor of the RF rectifier is kept OFF ($GPIO_1$=0V). The $GPIO_2$ signal driving the RF switch is used to implement an OOK modulation to the backscattered waveform that is horizontal polarized. This 'direct' OOK modulation is energy-efficient: the RF switch losses between incoming polar V waveform and polar H backscattered waveform are around 0.33 dB. In this mode, the RF switch is driven only when $GPIO_2$ is at 3.3 V, drawing a typical current of 1.8 mA during the high states of the OOK modulation for a short backscattering sequence. Having a polarization separation between the incoming WPT and the backscattered waveforms reduces the impact of clutter and cross-jamming effects.

iii) Backscattering Modulator (Co-Polarization) – In this mode, the incoming WPT signal received by the polar-V antenna is routed to the backscattering rectifier by maintaining the RF switch in its default state ($GPIO_2$ = 0 V), while $GPIO_1$ drives the rectifier MOSFET, toggling between 0 V (OFF) and 3.3 V (ON). Consequently, a co-polarization (polar V) backscattered waveform OOK modulated by $GPIO_1$ signal is generated. The backscattering rectifier was initially simulated in ADS software (by using nonlinear harmonic balance techniques) fabricated on FR4 substrate (thickness 0.8 mm, relative dielectric permittivity: 4.4, dielectric loss: 0.2) and then characterized with a 10 kΩ load resistor (emulating the PMU input impedance).

As shown in Fig. 2, the measured RF-to-DC power conversion efficiency (PCE) of the backscattering rectifier reaches 29% at −5 dBm in the 868 MHz ISM band. With the added fail-safe switch, the efficiency decreases by at most 5%, resulting in a peak of 25% at −5 dBm. For the targeted BWS Bluetooth Low Energy (BLE) application in this work over short ranges (a few meters), and considering the European regulatory limit of 33 dBm effective radiated power (ERP) at 868 MHz, the expected power at the rectifier input typically lies between −10 dBm and −5 dBm.

State-of-the-art works at similar frequencies, power levels and Schottky diodes report peak efficiencies ranging from 24% to 56%, generally achieved using optimized voltage doubler or multi-stage Schottky-based rectifiers [9], [10]. In contrast, this work employs a half-wave rectifier, with the backscattering functionality, measured with a fixed 10 kΩ load, which is not the optimal resistance. In practice, the PMU implements Maximum Power Point Tracking (MPPT), enabling dynamic load optimization and improved energy extraction.

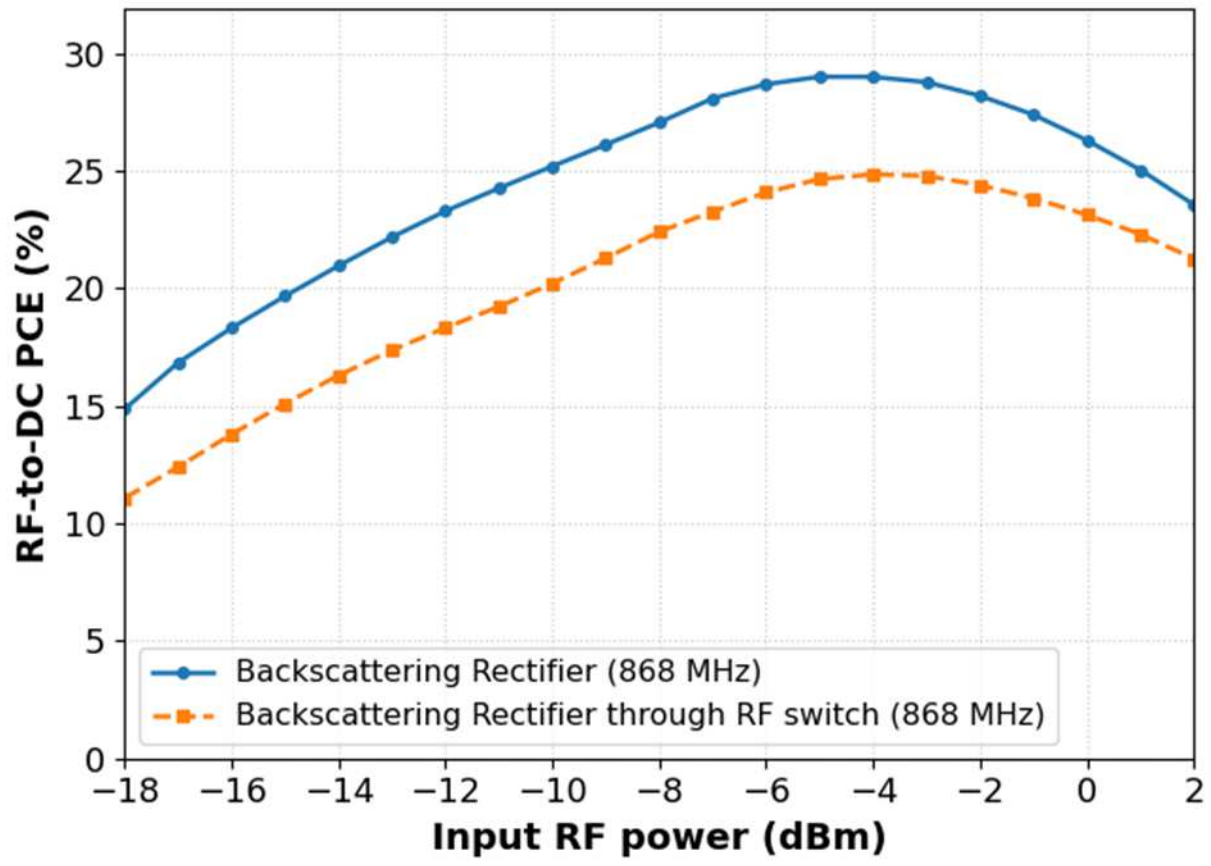

Fig. 2. Measured RF-to-DC Power Conversion Efficiency (PCE) at 868 MHz versus input RF power for the backscattering rectifier, compared with the backscattering rectifier including the fail-safe RF switch (RL=10 kΩ).

In a second stage the matching circuit of the rectifier was tuned with a real load that is the PMU of BLE based BWS. This on PCB tuning was necessary because the 10 kΩ load resistor not emulate exactly the PMU behavior and its associated MPPT algorithm.

The proposed BWS is composed of: the 2-bits BR, a power management unit (AEM3090, Epeas), a BLE Systems-on-Chip SoC (QN9080, NXP), a storage capacitor of 220 µF and a temperature/relative humidity sensor (HDC2080, Texas Instrument). All the system is able to start from an empty energy state of the storage capacitor. The PMU allow a cyclic operation between two pre-programed voltages threshold of the storage capacitor: minimal voltage $V_{min}$=1.2V and maximum voltage $V_{max}$=3.8V. The BLE SoC is shutdown during the capacitor charging process and activated by the PMU when $V_{max}$=3.8V. To save energy the BLE SoC operates in broadcaster mode sending the data in the advertising packets on three primary advertising channels (37, 38, and 39) without any incoming connection. On network level the data sent by the BLE BWS are received by a BLE hub (implemented with an NXP QN9080DK development card) operating as observer scanning continuously on passive mode, the advertising events on BLE channels. The BLE based BWS was selected for this proof-of-concept demonstrator because of its very low-power consumption [11]. When operating in an unpaired mode with basic cryptography, a BWS in BLE mode is vulnerable to various attacks, such as unauthorized intrusion, replay, relay, and eavesdropping attacks [12]. Adding a redundant and versatile backscattering channel, operating in the same ISM frequency band used for WPT, is a useful method to increase the security and resilience of BWS wireless communication against the aforementioned attacks by enabling identification through backscattering, agnostic to the underlying IoT protocol (BLE) [13].

## III. Experimental Results

To evaluate the proposed BWS, a dedicated setup in an indoor real environment was used as represented in Fig. 3a).

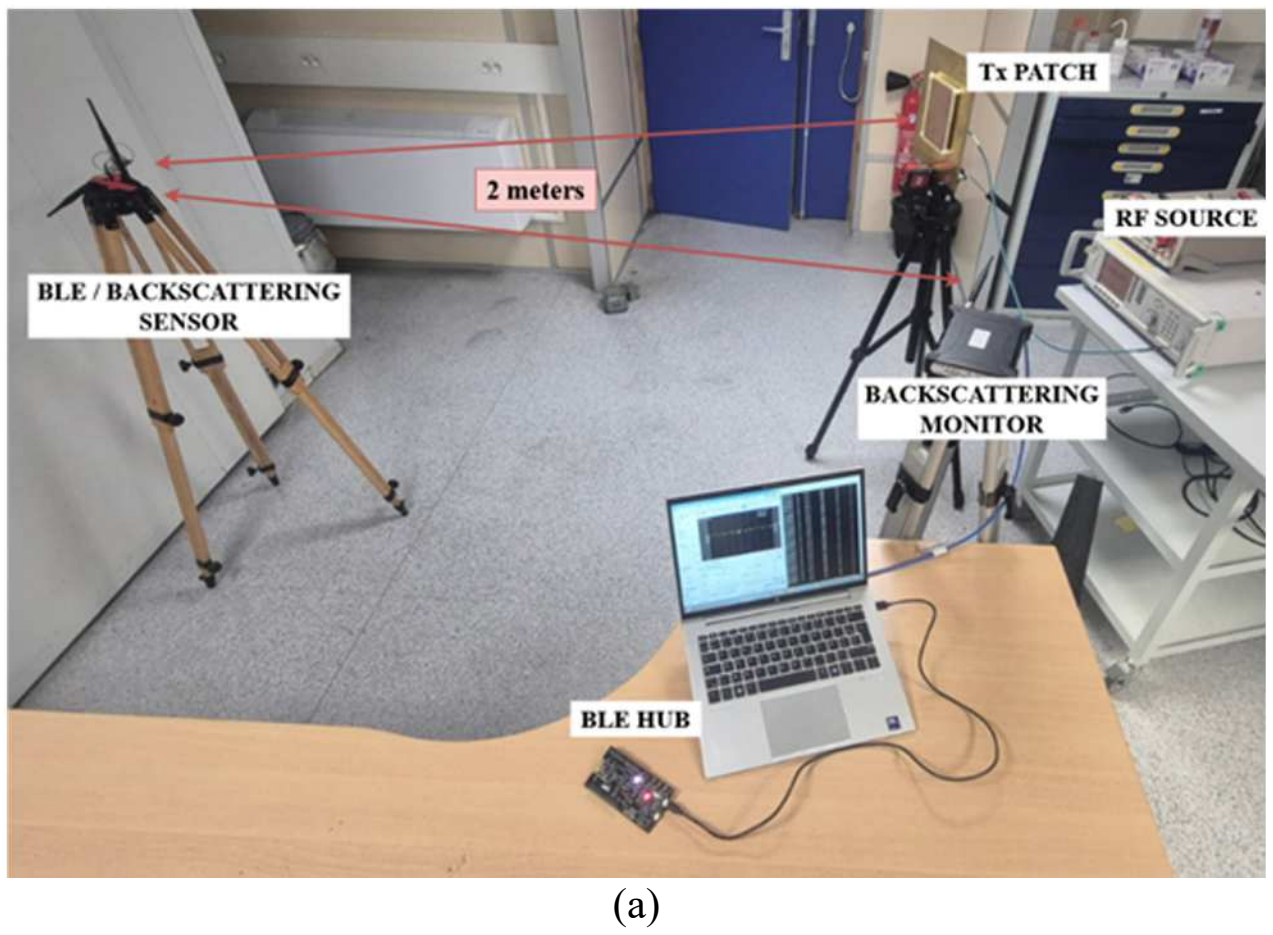

(a)

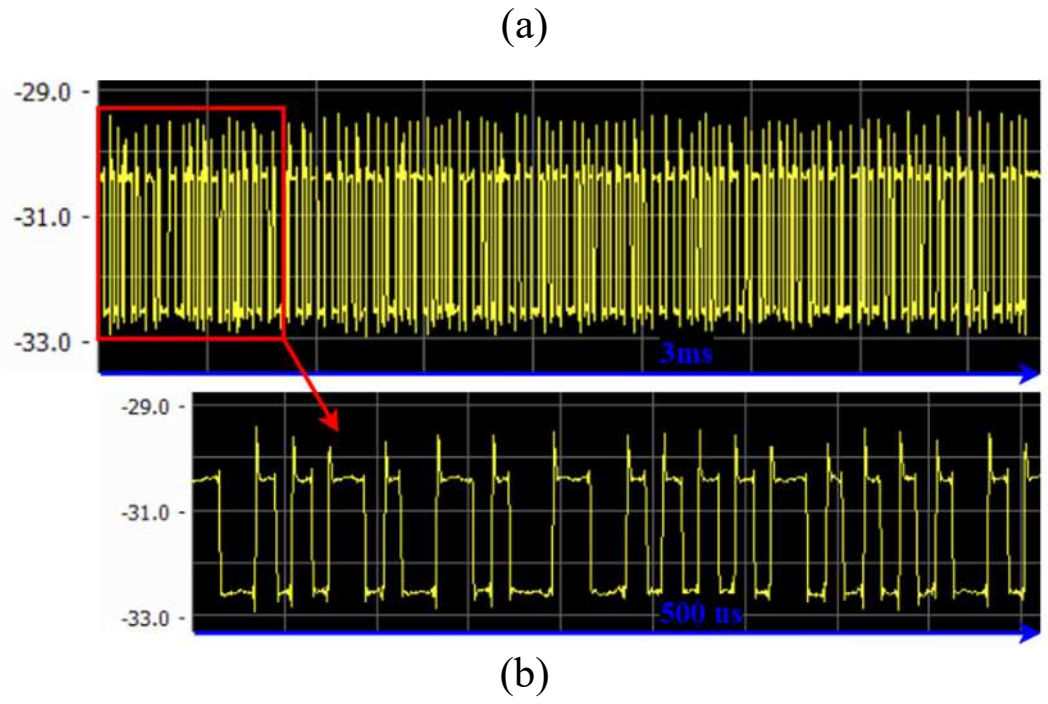

(b)

Fig. 3. (a) Experimental setup in a real indoor environment; (b) Backscattered cross-polarization frame (duration: 3 ms) captured by the real-time spectrum analyzer acting as a backscattering monitor, with a 500 µs time-window zoom.

The experimental setup consists of an RF signal generator (Anritsu MG3694B) delivering a 24 dBm WPT waveform at 868 MHz, connected to a vertically polarized home-made patch antenna (gain: +9.2 dBi) acting as the RF source. The backscattered signal is monitored using a Tektronix RSA306B USB real-time spectrum analyzer, connected to a horizontally polarized monopole antenna (gain: 2 dBi). The BWS is positioned at a distance of 2 meters from the transmitting antenna. A BLE hub is used to retrieve the BLE payload.

The BWS advertises data via BLE four times for redundancy on different channels. A 3 ms Manchester-encoded backscatter modulation 18-byte sequence is intercalated before each BLE advertising event. As shown in Fig. 3b), in both backscattering modes, the transmission is successfully received and captured by the spectrum analyzer.

Fig. 4 presents the complete backscattered frame with the decoded and annotated bytes. The frame consists of a 2-byte preamble for synchronization, followed by a 16-byte payload encrypted using AES-128. The payload includes 2 bytes corresponding to temperature and humidity measurements, interlaced with a 14-byte identification sequence. The resulting 16-byte AES-encoded block constitutes a Private Key (PvK), which acts as a dynamic private identifier of the BWS and is transmitted over the backscattering channel prior to each BLE advertising event.

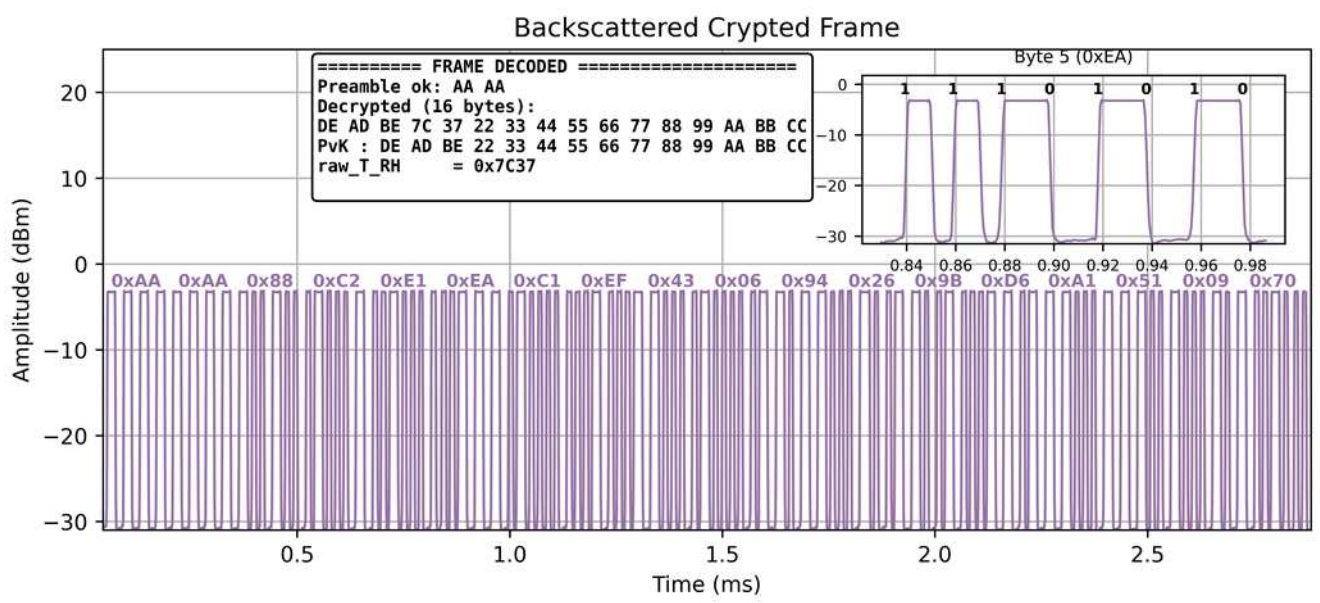

Fig. 4. Backscattered cross-polarization Manchester encoded and AES-128 encrypted frame with annotated decrypted values. This waveform was captured with the spectrum analyzer directly connected to the V-polar port of the BWS positioned 2 meters away from WPT source.

The captured backscattered frame was decrypted using a homemade Python script (not detailed here) with the corresponding AES-128 encryption credentials. The decrypted values are also reported in the inset of Fig. 4). Backscattering in co-polarization was also successfully captured and decrypted in the same manner (not reported here), demonstrating the capability of operating in two distinct backscattering modes and thus validating the functionality of the 2-bit rectenna. This enables the possibility of transmitting additional payload or identification data within the BWS.

Experimental results obtained in a real indoor environment demonstrated that, without applying any dedicated signal processing or correlation techniques at the backscattering monitor, a backscattering range of 2 meters was effectively tested in cross-polarization (polar H) and approximately 1.5 meters in co-polarization (polar V).

The energy overhead of the backscattering modes was quantified. In the original mode (without backscattering), the BWS operates in a periodic cycle including initialization, sensor reading, and BLE advertising. The backscattering modes insert a 3 ms GPIO-driven sequence immediately before each advertising event. Current profiles were measured for three cases: (i) without backscattering, (ii) with rectifier backscattering, and (iii) with RF switch backscattering. The resulting time-domain current traces (Fig. 5) allow a direct comparison of the overhead introduced by each configuration.

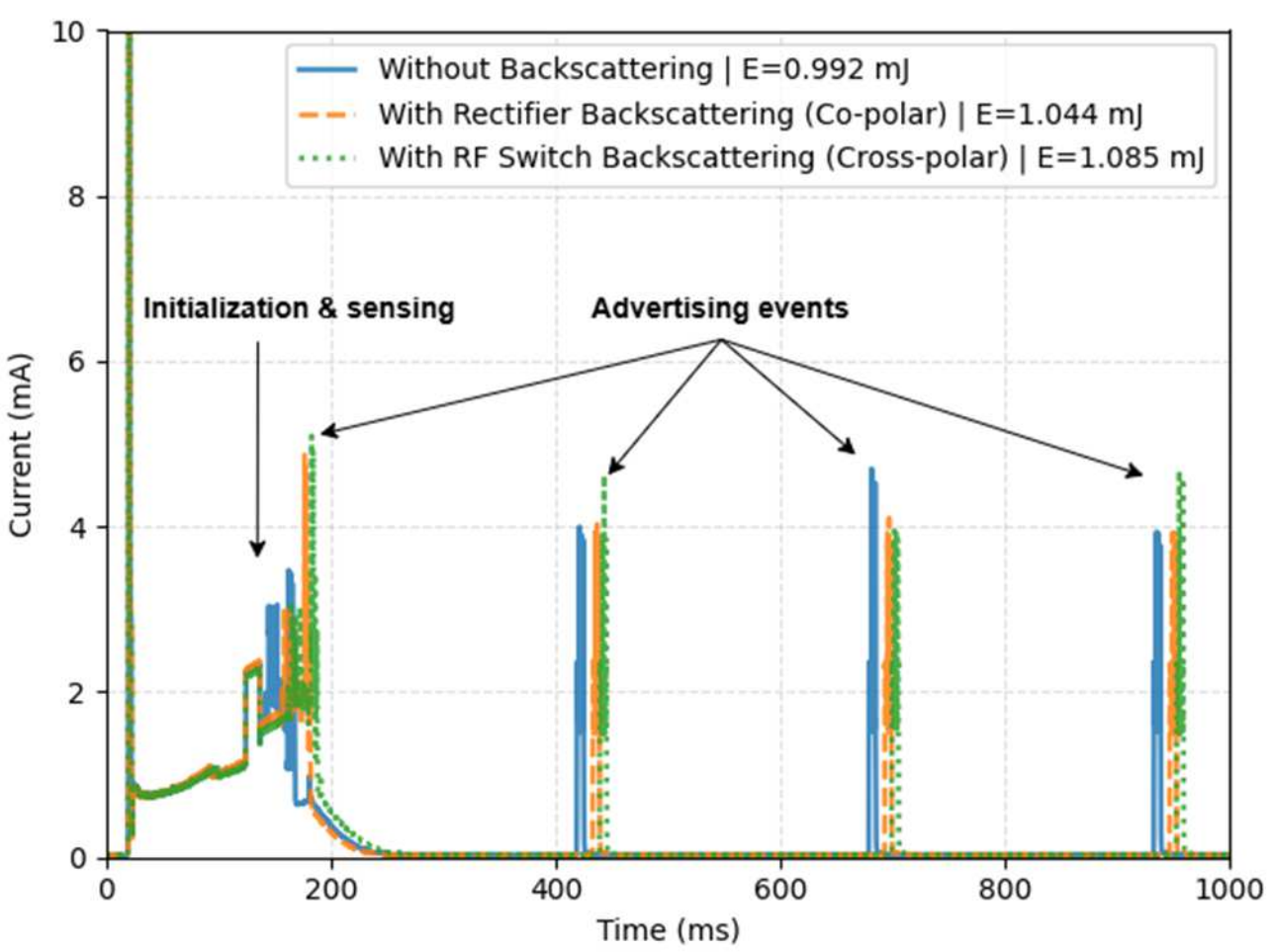

Fig. 5. Measured current consumption (3.3V bias) versus time over one complete BWS operating cycle, comparing the different modes (without backscattering, with rectifier backscattering, and with RF switch backscattering).

The insertion of the 3 ms backscattering sequence shifts the activity timeline in both configurations (co- and cross-polarization). The rectifier backscattering mode introduces an additional energy cost of 52 µJ, whereas the RF switch backscattering mode results in 93 µJ, corresponding to an extra 41 µJ compared to the rectifier-based mode. In both cases, the added energy remains marginal relative to the total energy consumed during one complete operating cycle.

## IV. Conclusion

This paper presents a new dual-polarized rectenna controlled by two digital bits and integrated into a battery-free wireless sensor (BWS) as a proof of concept. The rectenna operates either as an energy harvester or as a backscattering modulator using two orthogonal polarizations (co- and cross-polarized), enabling flexible mode switching through two digital control signals. Using this versatile rectenna within a BLE-enabled battery-free platform, a 16-byte AES-128 encrypted payload was transmitted via co- and cross-polarized OOK-modulated backscattered waveforms, combining wireless power transfer, dual-polarized backscattering, and secure payload transmission in a compact system. This dual-mode capability enhances communication redundancy and improves resilience against replay, relay, and eavesdropping attacks. Future work will focus on deeper electromagnetic characterization, including comparative RCS analysis, polarization isolation evaluation, and further optimization of backscattering efficiency and range.

## Acknowledgment

This research was funded, in whole or in part, by the French National Research Agency (ANR) under the project SWAVE "ANR-25-CE39-5853-01", and the authors gratefully acknowledge its support.